# An statistical analysis of stratification and inequity in the income distribution

Juan C. Ferrero[a]

*Centro Laser de Ciencias Moleculares, INFIQC – Departamento de Fisicoquímica, Facultad de Ciencias Químicas. Universidad Nacional de Córdoba - 5000 Córdoba - Argentina*

Short title: Stratification and inequity in the income distribution

The analysis of the USA 2001 income distribution shows that it can be described by at least two main components, which obey the generalized Tsallis statistics with different values of the *q* parameter. Theoretical calculations using the gas kinetics model with a distributed saving propensity factor and two ensembles reproduce the empirical data and provide further information on the structure of the distribution, which shows a clear stratification. This stratification is amenable to different interpretations, which are analyzed. The distribution function is invariant with the average individual income, which implies that the inequity of the distribution cannot be modified by increasing the total income.

[a] jferrero@fcq.unc.edu.ar

# 1 Introduction

The individual income distribution provides the basic information for the analysis of the social situation and the results of specific economical policies[1]. The main question it poses refers to the origin of its shape, the inequity associated with it and the stratification of the society. The first of these points, the shape, presents an intriguing aspect when the empirical data are viewed on a log-log cumulative plot, and although it has received much attention since it was first noted by Pareto [2], only a partial explanation has so far been attained. Such a plot shows a concave curvature in the region that corresponds to agents with low and medium income and a linear behavior at high values. The transition between both regions is not smooth but, on the contrary, it is characterized by an abrupt change of slope. This discontinuity lead to represent the whole distribution by two separated functions, each one describing a different income region, thus reflecting a stratification of the distribution, without apparent connection between them. At present it is well established that the high income limit, the Pareto region, which comprises a small and extremely wealthy group of agents, follows a power law of universal character and, in spite of some fluctuations, temporal stability [3]. On the contrary, the lower part of the distribution has received less attention, even though it represents an ample majority of the population, typically more than 90%. Attempts have been made to fit this part to various functions, such as the usual exponential, gamma and log-normal distributions [4-8] in addition to the more recent Tsallis [9] and the Kanadiakis [10,11] generalized distributions.

On the theoretical side many efforts have been made to explain the empirical behavior and they have been recently reviewed [6, 12]. One widely used approach has been treating the individual income as a stochastic process from which the distribution arises [13-15]. A new approach was also proposed, which applies the methods of the statistical mechanics to develop a multiagent model, in a closed economy [16-21]. One of these model, isomorphic with the others, has been first independently developed and applied by Angle in a series of paper [17, 18]. From the computational point of view, this model is a particular case of the more general model but as they differ on the basic premises, their properties are different. These models have the common feature of treating the economic agents as stochastic scattering particles that interchange a positive amount of money, like the ideal gas. After a number of these exchanges, a stationary distribution is attained. In the earlier studies with this model the stationary distribution obtained corresponded to a single exponential decay, in agreement with the Gibbs – Bolztmann distribution of energy of an ideal monatomic gas [19, 20]. It was soon evident that this exponential decay could not reproduce the whole range of empirical data. Therefore, the model was later modified to include the ability of the economic agents to save part of the money [21, 22]. This propensity saving factor successfully predicted the general qualitative profile of the distribution: a Gamma-like shape at low income and a Pareto tail in the high income limit, but with a smooth transition between them.

These models have been recently comparatively [22] and critically analyzed [23].

The main objective of most of the theoretical studies has been to reproduce the Pareto tail and to qualitatively account for the presence of a maximum in the probability density function (PDF). Model calculation using two groups of agents with different saving propensities that qualitatively accounts for the general features of the distribution has also been reported [24-26]. However, little progress has been made to quantitatively explain the empirical behavior of the entire distribution.

In addition, the difficulty in the analysis of the empirical data is not minor, as have been recently pointed out [23,27]. Thus, the cumulative distribution function (CDF) is rather insensitive to the details of the distribution while the PDF is too dependent on the selection of the bin size. Therefore, extreme care should be exercised in processing the empirical information, which should also be of high quality and provided by a very dependable source. In this respect, two very complete and consistent pieces of data, available over an extended period, are those provided by the official statistical agencies of Japan and the USA. In this work we will concentrate on the data from the USA, fiscal year 2001 (USA2001), since although the probability distribution changes with time, the general features, which are of importance for the theoretical analysis, remain.

The purpose of this work is to obtain adequate fits to the empirical data, to analyze the income distribution to the light of the gas kinetic model and, relying on these results, to obtain clues to the origin of its shape, composition, social stratification and inequity.

The results show that the population can be broadly divided in two groups of agents that originate the observed stratification. One group, associated with additive processes and agents whose sources of income are mainly wages and salaries, contributes to the low and medium income region. The other group, which appears in the high income region is actually part of a broader distribution, from low to very high income and corresponds to a multiplicative process, typical of investors.

## 2  Analysis of the empirical data

This work is constrained to the USA data corresponding to fiscal year 2001, as representative of the general behavior. Two pieces of information are relevant to the present work, the CDF, that can be directly obtained from the pertinent Internal Revenue Sservice (IRS) table and the PDF which is easily calculated from the returns as a function of income size, also provided by the IRS. These data are plotted in Figures 1 and 2 respectively.

The data were fit to the probability function derived by Tsallis in his formulation of a generalized Statistical Mechanics, which has been successful in describing the behavior of non-extensive systems [28,29]. This function, hereafter denoted by $Ts(x)$, is:

$$Ts(x) = Nh(x)[1 - (1-q)\beta^* x]^{1/(1-q)} \qquad (1)$$

where $\beta^*$ is the Lagrange parameter and the value of $q$ depends on the system under study. The factor $h(x)$ represents, in the physical counterpart, the degeneracy of states. This expression is a generalized distribution which includes the usual Gibbs-Boltzmann function $B(x)$, as a special case, in the limit $q \to 1$:

$$B(x) = g(x) e^{-x/\beta} \qquad (2)$$

Here $g(x)$ replaces $h(x)$ and $\beta$ is the usual Lagrange constraint ($\beta = 1/\beta^*$).

In the absence of a knowledge of the functional form of $h(x)$, it could be represented as a power series in the variable $x$. However, considering the precision of the data, in the present context it is sufficient and convenient to keep only one general term in the expansion, as an useful approximation:

$$h(x) = K x^n \qquad (3)$$

where $n$ is a real number. In this case the Tsallis function in the limit $x \to \infty$ reduces to:

$$\lim_{x \to \infty} P(x) = N[(1-q)\beta^*]^{1/(1-q)} x^{n + 1/(1-q)} \qquad (4)$$

If $q > 1$ and $n \leq 1/(q-1)$ the exponent $n + 1/(1-q)$ is negative and this equation produces the inverse power law expected for a Pareto behavior.

The empirical data for the QDF and PDF were fit according to the following procedure. By definition, the QDF, $Q(x)$, is obtained from integration of the PDF

$$Q(x) = \int_x^\infty P(x)dx \tag{5}$$

$P(x)$ was taken as the sum of a Bolztmann-like function, $B(x)$ (That is, a Tsallis function with $q = 1$) and a Tsallis function with $q > 1$, $Ts(x)$, with relative weights $N$ and $M$, respectively:

$$P(x) = NB(x) + MT(x) \tag{6}$$

In the present work satisfactory results were obtained using $h(x)$ in the Tsallis term as

$$h(x) = x^\varphi \tag{7}$$

and for $g(x)$ the sum of two terms:

$$g(x) = Ax^\alpha + Cx^\kappa \tag{8}$$

where $\alpha$ and $\kappa$ are real adjustable parameters. With this choice Eq.2 becomes the addition of two Gamma functions with the same value of $\beta$ and mean values $(\alpha+1)x$ and $(\kappa+1)x$. Calculations with only one term for $g(x)$ did not produce satisfactory results.

In the fitting process, $P(x)$ was calculated by optimization of the parameters of Eqs. 1, 2 and 6 to yield the following normalized expressions for $B(x)$ and $Ts(x)$:

$$B(x) = (1.46x^{0.133} + 3.65x^{1.63})\exp(-x/0.406) \tag{9}$$

$$Ts(x) = 3.03(1 + 0.90x)^{-3.45} x^{0.88} \tag{10}$$

and $\quad P(x) = 0.9B(x) + 0.1Ts(x) \tag{11}$

These equations were subsequently numerically integrated to obtain *Q(x)*. The results of the fits are shown in Figures 1 and 2. It should be noted that the set of parameters presented in Eqs 9-11 are not unique. Considering the complexity of the functions involved and the uncertainty of the empirical data, other sets also produce satisfactory fits. The parameters, however, do not show a variation that could affect the analysis of the data, at least in the present context.

A detailed analysis of these results will be made after presenting the results obtained from the model calculations. For now, the pertinent evidence is that these results unambiguously show the multicomponent character of the income distribution, hereafter designed as B and T, corresponding to the Gibbs-Boltzmann and the Tsallis distribution, respectively, with group B in fact composed of two subgroups.

## 3 Model calculations

To obtain further insight on the income distribution, the empirical data were simulated according to the theoretical model developed by the Kolkata School [20]. This is a multiagent model of a closed economy, so that the number of agents and the total amount of money are constant. Agents are allowed to interact stochastically and at every time step two of them, *i* and *j*, randomly exchange a certain amount of money, *m*, so that

$$m_i(t') + m_j(t') = m_i(t) + m_j(t) \tag{12}$$

or, in other terms, they exchange an amount of money *Δm*, so that

$$m_i(t') = m_i(t) + \Delta m \tag{13}$$

and

$$m_j(t') = m_j(t) - \Delta m \tag{14}$$

where the value of $\Delta m$ is calculated according to a prescription that constitutes the trading rule between the agents.

The simplest version of the trading rule calculates the value of money of each agent after interchange as a random fraction $\varepsilon$ of the total amount of on money involved in that exchange:

$$m_i(t') = \varepsilon[m_i(t) + m_j(t)] \tag{15}$$

$$m_j(t') = (1-\varepsilon)[m_i(t) + m_j(t)] \tag{16}$$

so that

$$\Delta m = \varepsilon[m_i(t) + m_j(t)] - m_j(t) \tag{17}$$

This simple model results in an single exponential distribution of income, similar to that for the energy of a monatomic ideal gas, that therefore departs from the empirical observations. However, satisfactory results for both the low and medium income region as the high income part of the distribution are obtained when the trading rule incorporates a saving propensity factor, $\lambda$. for each agent. In any trading, the agents save a fraction $\lambda$ so that the trading rule becomes

$$\Delta m = \varepsilon(1-\lambda_j)m_j - (1-\varepsilon)(1-\lambda_i)m_i \tag{18}$$

For constant $\lambda \neq 0$, the steady state distribution of money decays exponentially on both sides of the PDF with a mode locate at values that increase with $\lambda$. Distributed values of $\lambda$ results in a fat tail, that follows the power law expected for a Pareto behavior. The model is therefore able to qualitatively account for the main features of the empirical PDF through the impressive effect of savings. The main drawback is that it results in a smooth transition between the low and high income region, contrary to the empirical information.

The analysis of the empirical data presented in Section 2 shows that the distribution consists of two main components, each one with a different statistical behavior, as reflected by the values of *q*. Therefore, those two groups of agents must be introduced into the model. In the present calculations a total population of 1000 agents was divided in two sets B and T, with relative populations of 0.90 and 0.10 and they were allowed to interact through $t = 10^6$ trades. The steady state distribution was calculated as the average of the results for 200 initial configurations. Neither a fixed constant value of $\lambda$ nor a uniform distribution produced a satisfactory fit to the cumulative distribution of the USA for year 2001. Instead, calculations with $\lambda_B = 0$ and $\lambda_T$ given by a distribution of quenched values [21]

$$\lambda_{Ti} = 1 - \varepsilon_i^{1/\alpha} \qquad (19)$$

with $\alpha = 1.25$ produced the results shown in Figs.1 and 2 for the total population and its two main components

## 4 Discussion

### 4.1 Stratification

The analysis of the empirical data and the model calculation results indicates that the income distribution has a stratified structure, which can be considered from various aspects.

The most obvious stratification is based on income. On these grounds two different situations arise, depending of whether the income range or the income average is considered.

In the particular case studied, there is a transition point located at $x \approx 4$ that divides the distribution in two regions ( Fig. 1). The region located at $x > 4$, here after symbolized by P, is characterized by a prevailing Pareto behavior and it is almost exclusively composed of the tail of the distribution of those agents whose behavior is described by the Tsallis function with $q = 1.28$ (group $T_P$) with a negligible contribution from agents whose income follows the gamma PDF (group $B_P$). The other region, non-Pareto, indicated as NP, is situated at $x < 4$, and consists of the dominant group $B_{NP}$, which obeys a Boltzmann statistics, plus the group of agents that follows Tsallis statistics in this region, $T_{NP}$. Consequently, the Pareto and the non-Pareto regions constitute the two most obvious strata that arise on consideration of the income range of the distribution.

A different stratification appears when the specific income of group $T_{NP}$ is considered. Integration of Eqs. 9 – 11 yields the fractional population while integration of the same equations times *x* provides the corresponding amount of money. The calculations show that even though in the non-Pareto region the population of group $B_{NP}$ outnumbers that of group $T_{NP}$ by a factor of 12, the average income of $T_{NP}$ agents is larger by a factor 1.86 on a per agent basis. Considering that $B_P$ is negligible, so that $B \approx B_{NP}$, the difference in individual average income allows for a stratification in three groups, in increasing order of income per agent: group B , group $T_{NP}$ and group $T_P$, with average incomes of 0.76, 1.86 and 52, respectively.

The third stratification arises from the different statistics followed by the agents, either Gibbs-Boltzmann or Tsallis. In this case, the Pareto law that describes the income distribution that corresponds to hyper-rich agents (group $T_P$), is just the high income limit of group $T = T_{NP} + T_P$, whose income encompasses the whole range of values, starting from zero. These agents have access to very large individual income, but this by itself does not lead to economic success. In fact, nearly 80% of them falls in the low-middle income group ($T_{NP}$) . However, as mentioned above, the average income of group $T_{NP}$ is larger than that of B, and therefore, the income of individuals belonging to group T always is larger that those of group B,  which, in addition, never reaches the high income region.

An insight of the origin of stratification can be obtained from the model calculations. The property that characterizes each agent is the saving propensity. Therefore, groups B and T are defined by the different attitude towards money, through the value of $\lambda$. The theoretical results indicate (Figure 3) that group $T_{NP}$ consists of agents with low values of the saving propensity factor, while the opposite holds for those in the Pareto tail ($T_P$).

It has been suggested that this dual behavior arises from the different activities on which income relies [24].  The power law behavior corresponds to agents with income based in multiplicative processes, that is, the Pareto region belongs to the realm of  investors and entrepreneurs, while the low-middle range, that follows a Gibbs-Bolztmann statistics is composed of agents whose main income originates in wages and salaries, an additive process.

The Tsallis distribution in the non-Pareto region could also be satisfactorily fit by a Gamma function, although the deviation rapidly increases with $x$ above 4, indicating the tendency of agents in group $T_{NP}$ to follow an additive processes.

Interestingly, the income distribution of agents in group *B* (the employees) seems to show two components with widely different average income values, 0.46 and 1.07.  This implies an additional stratification of the society.  There is strong evidence, based on US IRS data as well as statistics from other countries that the apparent GB distribution has several components, each one corresponding to a different educational level [30]. The observation of only two groups in this study is only indicative of the difficulties in obtaining detailed information from macroscopic statistics, as it happens for physical systems.

**4.2 The inequity of the distribution**

Two widely used measurements of the inequality of the income distribution are the Gini coefficient and the ratio between the top 10% of the distribution to the bottom 10%, $R$. For simplicity, the following discussion will be focussed on the dependence of $R$ on the parameters that characterize the shape of the income distribution.

For a society with agents aggregated in $i$ components the income distribution can be expressed, in general terms, as a combination of Tsallis and GB (or Tsallis functions with $q=1$) functions, as is the case of USA 2001.

One important characteristic of the gamma and the Tsallis distributions is that, aside from the $q$ factor, they are characterized by two parameters, $n$ and $\beta$. While $n$, the exponent of $x$ in the degeneracy factor, is a shape parameter, $\beta$ ( or $\beta^*$) is a scale parameter, so that both functions are invariant on it. For a system in equilibrium there should be a unique value of $\beta$ for all the component groups and consequently, the ratio $R$ should not depend on it.

The mean value of the income distribution for the generalized gamma function is given by

$$\langle x \rangle = (n+1)/\beta^*[1+(n+2)(1-q)] = (n+1)\beta[1+(n+2)(1-q)] \qquad (20)$$

If the distribution corresponds to the ordinary gamma function, then $q = 1$ and Eq. 20 becomes

$$\langle x \rangle = (n+1)/\beta^* = (n+1)\beta \qquad (21)$$

Eqs 20 and 21 hold for a single component. In a many components society the mean value is

$$\langle x \rangle = \sum c_i \langle x_i \rangle \qquad (22)$$

where $c_i$ is the relative weight of the $i^{th}$ group. In a set of groups in equilibrium, all of them have the same value of β and consequently, <x> will be, in general

$$\langle x \rangle = \beta \sum c_i (n_i + 1)[1 + (n_i + 2)(1 - q_i)] = \beta \sum c_i (n_i + 1)k_i = \beta \sum c_i \alpha_i k_i \qquad (23)$$

where $k_i$ is a function of $n_i$ and $q_i$

For a single gamma function, $q_i = 1$, $k_i = 0$ and Eq. 22 reduces to

$$\langle x \rangle = \beta \sum c_i (n_i + 1) \qquad (24)$$

Consequently, the total income in a society increases with *n* and with β. For constant $n_i$ the total richness will increase with β but since the distribution function is invariant on it, the ratio *R* will not change. In the case of gamma functions, this result also holds for the Gini coefficient. The immediate conclusion is that increasing the GNP will not per se modify the inequity of an income distribution, although it will certainly produce a richer society. The inequity arises from the value of *n* and it is only modifying this shape parameter than a more equal distribution can be obtained.

This analysis also pertains to the evolution of the distribution in a complex system when the agents are in internal equilibrium within the group to where they belong, but not in equilibrium with the rest of the groups that constitute the total system. Hence, each group has a different initial value of β and the original population evolves towards a stationary distribution characterized by a unique final value of β for every group in the ensemble.

On the premise that the total population and the total amount on money is constant, the mean value of money should be time invariant

$$\langle x \rangle_i = \langle x \rangle_f \qquad (25)$$

$$\sum_i^n c_i \alpha_i k_i \beta_i^i = \sum_i^n c_i \alpha_i k_i \beta_i^f = \beta^f \sum_i^n c_i \alpha_i k_i \qquad (26)$$

and

$$\Delta x_j = \langle x_j \rangle_f - \langle x_j \rangle_i = \alpha_j k_j \left( \beta^f - \beta_i^i \right) = \alpha_j k_j \frac{\sum_i^n c_i \alpha_i k_i \left( \beta_i^i - \beta_j^i \right)}{\sum_i^n c_i \alpha_i k_i} \qquad (27)$$

Therefore if $\left( \beta_i^i - \beta_j^i \right) > 0$, $\Delta x_j$ will be also be positive. This means that in a nonequilibirum distribution, money will flow from those groups with higher β to that with a lower value, irrespective of the initial mean value of money, until a stationary state, with a single value of β is reached.. In other words, richness (or poorness) does not provide a criterion to predict the flow of money, but it is evolution to a state with an unique value of β which determines the direction of the change of the distribution. After the stationary state is reached, the inequity of the income distribution will be determined by the value of α for each group of the system.

In principle, these considerations apply to any human ensemble, either the people of a country or even the countries of the world. In the latter case, it could provide grounds to analyze the consequences of globalization, within the limitations imposed by the small number of countries in the world.

## 5 Conclusions

An analysis of the empirical income distribution of USA 2001 (both the PDF and the QDF), shows that it consists of at least two well differentiated components, which are well described by Tsallis functions with different values of the parameter *q*, and with the appropriate degeneracy factors. One of these functions has $q = 1.28$ and in the high income limit shows the expected Pareto behaviour. The other term has $q = 1$, and, as a consequence, follows the usual Gibbs-Boltzmann statistics, which actually is a Gamma function. Therefore, the complete PDF is described by the addition of two terms, with coefficient 0.90 for the Boltzmann component and 0.1 for the Tsallis function.

The empirical data could be reproduced using the gas kinetics model. In these calculations the total population consists of two sub-ensembles: one with a distributed saving propensity factor and the other without savings and relative populations of 0.1 and 0.9, respectively. The Pareto tail consists almost exclusively of agents belonging to the first group, while the low-medium income region presents both components with predominance of the latter. The income distribution of the agents with a distributed saving propensity factor follows a Tsallis statistics with $q = 1.28$, while the group with $\lambda = 0.$ is represented by a Gamma distribution. In addition, within the first group, those agents with the largest saving factor contribute mainly to the Pareto tail. Therefore, the Pareto tail is just the minor visible component of a sub ensemble of the system, whose income distribution follows Tsallis statistics and that is mostly embedded into the Gamma distribution of the main group.

The results provide the basis to analyze the social stratification in different ways and it is also indicative of the influence of education on the income distribution, as noted form empirical data from the USA and other countries.

A direct consequence of the shape of the income distribution and the function that produces a fit to it is its invariance on the average income. This simple means that increasing the total income, as measured for instance by the gross national product, although it certainly implies a richer society, does not result is a change of the shape of the distribution. Therefore, inequity relies on the value of the exponent in the degeneracy factor, which seems to depend on the educational level.

**Acknowledgements**. The author thanks CONICET and FONCYT for financial support.

**Captions to the Figures**

**Figure 1.** Cumulative distribution function as a function of money, in relative units: (■) Empirical data ; ( ----) Fit to the empirical data; (○) Model calculations: component with $\lambda = 0$ and (▬) the corresponding fit to a gamma function; (○) Model calculations: component with distributed $\lambda$ and (▬) the corresponding fit to the Tsallis distribution. The total for the model calculations is the addition of the parcial components and it is not shown, for clarity ( See, however, Figure 2).

**Figure 2.** Probability density distribution as a function of money, in relative units: (■) Empirical data ; ( ----) Fit to the empirical data (overlapped with the other curve); (○) Model calculations: component with $\lambda = 0$ and ( ▬) the corresponding fit to a gamma function; (○) Model calculations: component with distributed $\lambda$ and (▬) the corresponding fit to the Tsallis distribution. ( ▬) Fit to the total PDF for the model calculations as the addition of the partial components.

**Figure 3.** (■) Contribution to income, in relative values, as a function of the saving propensity factor, form the model calculations. The solid red line is only a guide to the eye.

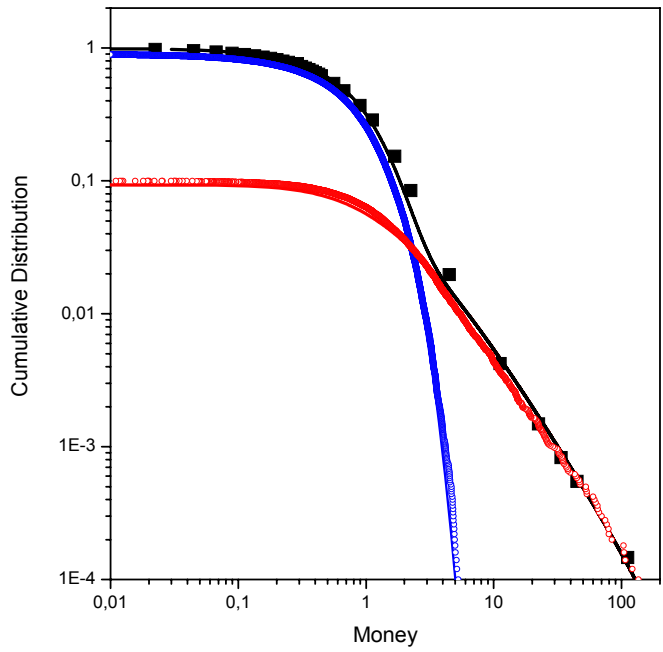

Figure 1

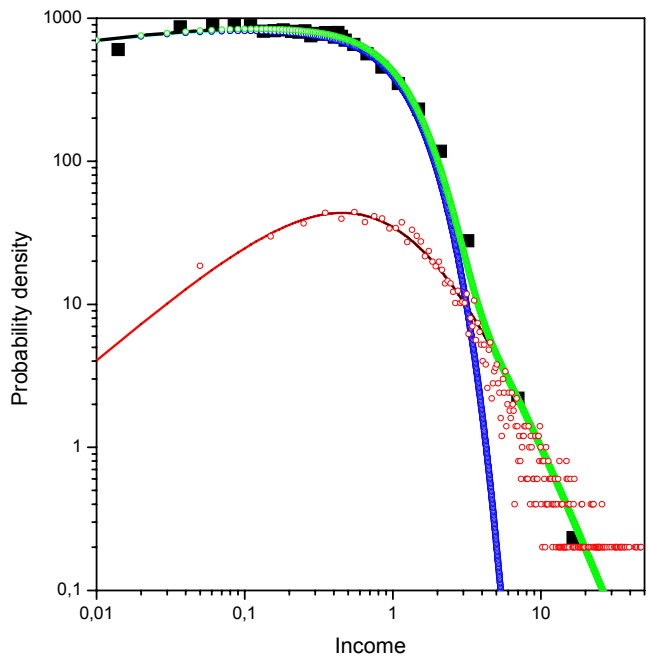

Figure 2

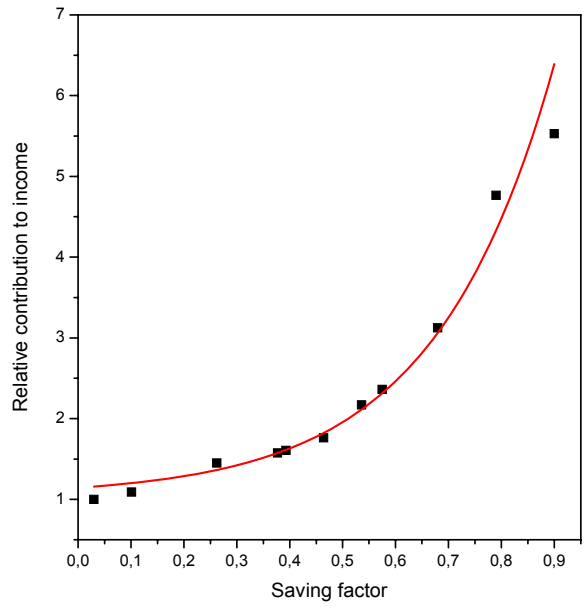

Figure 3.